\documentclass[12pt,preprint]{aastex}

\shorttitle{Radio Emission from a Fast Halo CME}
\shortauthors{Bastian}

\begin{document}

\title{Synchrotron Radio Emission from a Fast Halo Coronal Mass Ejection}
\author{T. S. Bastian}
\affil{National Radio Astronomy Observatory\footnote{The National Radio Astronomy Observatory is a facility of the National Science Foundation operated under cooperative agreement by Associated Universities, Inc.}, Charlottesville, VA 22903}

\begin{abstract}
\noindent An interplanetary (IP) type-II-like radio burst is analyzed. It occurred on 2003 June 17-18 in association with a fast halo coronal mass ejection (CME), an M6.8 soft-X-ray (SXR) flare, and produced a solar proton event. Unlike coronal type II bursts and the majority of IP type II radio emissions, the IP type-II-like event associated with the fast halo CME on June 17-18 varies smoothly in time and frequency and has a frequency bandwidth that is several times larger than is typical for coronal and IP type II emissions. Moreover, the frequency change with time is inconsistent with that expected from plasma radiation associated with a CME-driven shock. I suggest that this IP type-II-like event, referred to here as an IP type II-S event, is not due to plasma radiation but, rather, incoherent synchrotron radiation from near-relativistic electrons entrained in the CME magnetic field, or in the sheath region between the shock and the CME driver. This event may be an example of a new and distinct class of interplanetary radio phenomenon. 
\end{abstract}

\keywords{Sun: coronal mass ejections --- Sun: flares --- Sun: radio radiation --- Sun: X-rays}

\section{Introduction}

Coronal and interplanetary type II radio emissions are widely believed to result from the interaction of a fast shock with the coronal and/or the interplanetary medium (Wild 1950, Malitson et al. 1973, Boischot et al. 1980, Nelson \& Melrose 1985). The sequence of events is thought to be as follows: the shock accelerates a distribution of electrons to suprathermal ($\sim\!10$ keV) energies; the electron energy distribution is unstable to the production of Langmuir waves; the Langmuir waves, in turn, are converted by means of nonlinear wave-wave interations to electromagnetic waves -- plasma radiation -- near the fundamental and/or harmonic of the local electron plasma frequency. In the case of coronal type II radio bursts, the nature of the shock driver remains controversial: flare blast waves (Uchida 1974, Wagner \& MacQueen 1983; Cane \& Reames 1988; Vrsnak et al. 2006), flare ejecta (Gopalswamy et al. 1997), and CMEs (Cliver et al. 1999, 2004) have all been considered. In contrast, it is generally accepted that shocks driven by fast CMEs are responsible for IP type II radio emission (Cane 1984, 1985; see also the recent review by Gopalswamy 2006). Regardless of the driver, type II radio emission has long been used as a tracer of shocks in the corona and the interplanetary medium. Since plasma radiation is emitted at frequencies near the electron plasma frequency $\nu=\nu_{pe}$ and/or $\nu=2\nu_{pe}$, where $\nu_{pe}=(e^2 n_e/\pi m_e)^{1/2} \approx 9 n_e^{1/2}$ kHz, it provides a direct link between the observed radiation and the electron number density in the source. Moreover, as the shock propagates through the coronal or interplanetary medium, the change in the frequency of the type II emission with time jointly constrains the density gradient and the shock speed along its trajectory. Observations of coronal and interplanetary type II radio bursts have been important in connecting phenomena observed in the corona (e.g., flares and CMEs) to {\sl in situ} measurements made near 1 AU (e.g., Leblanc et al. 2001, Reiner et al. 2001).

Cairns et al. (2000) posed the question of whether multiple classes of interplanetary type II radio events exist, drawing a distinction between ``broadband, smooth events" (Cane et al. 1982, Lengyel-Frey \& Stone 1989) and ``narrowband, intermittent events" (e.g., Reiner et al. 1997, 1998). Recently, Cane \& Erickson (2005) presented a study of events observed by the space-based WIND/WAVES experiment and ground-based spectrometers. Based on a qualitative evaluation of 1-14 MHz spectra, they  conclude that there are indeed multiple classes of type II radio emission and classify them into three groups: 1) extensions of coronal type II radio bursts from meter wavelengths into the 1-14 MHz frequency range; ii) ``blobs and bands"; iii) ``IP type II events". Their classifications are non-exclusive -- more than one class of type II phenomenon can occur during a given event. Drawing from a sample of 135 type II burst events identified in the 1-14 MHz radio band from 2001--2003, Cane \& Erickson found that 70\% displayed ``blobs and bands" and $50\%$ showed evidence for the presence of extensions of coronal type II bursts. Less than $25\%$ of the events showed the presence of an ``IP type II event".  These events are therefore rare, occurring at a rate of approximate one per month during the years in question. 

On the other hand, Gopalswamy (2006) has suggested a unified approach to understanding the type II phenomenon. He emphasizes the hierarchical relationship between CME kinetic energy and the wavelength range and, hence, radial range over which type II emission occurs, and notes the ``universal" relationship between type II drift rates and their emission frequency (Vrsnak et al. 2001, Aguilar-Rodriguez et al. 2005a), as well as a ``universal" value for the frequency bandwidth ratio $\delta\nu/\nu$ (Aguilar-Rodriguez et al 2005b).

In the present case study a type-II-like event that was observed on 2003 June 17-18 is analyzed in detail. In fact, the event matches the definition of an ``IP type II event" and its initial developments is used as an example by Cane \& Erickson (2005; see their Fig.~9). ``IP type II events" are defined by Cane \& Erickson as those that i) start at a frequency $>1$ MHz but extend to frequencies $<1$ MHz; ii) have a duration $>3$ hrs; iii) are broadband ($\Delta\nu/\nu\sim 1$). On the basis of its observed properties, however, I show that it is difficult to reconcile with existing ideas regarding coronal and IP type II radio emission. In particular, I show that synchrotron radiation, rather than plasma radiation, may be the relevant emission mechanism. In other words, it may be necessary to distinguish between two, fundamentally different types of slow-drift radio emission in the interplanetary medium: the familiar {\sl IP type II-P emission} due to plasma radiation, and {\sl IP type II-S emission} due to synchrotron radiation. Since synchrotron radiation involves the interaction of relativistic electrons with magnetic fields, the phenomenon raises a number of critical questions regarding electron acceleration and transport. It also offers a new diagnostic tool for probing the nature of fast CMEs in the interplanetary medium (IPM). The observations are presented in \S2; problems with an interpretation of the data based on plasma radiation are pointed out in \S3; a simple synchrotron model is presented in \S4. I briefly explore some of the implications of this idea in \S5.

\section{Observations and Results}

The radio event of interest occurred on 2003 June 17-18,  type-II-like radio emission associated with a fast halo CME (Fig.~1), a soft-X-ray (SXR) flare of GOES class M6.8, and a solar proton event. The flare occurred in NOAA active region number 10386 at a position E55, S07.  

The radio observations were made by the WAVES experiment (Bougeret et al. 1995) on board the WIND satellite, a spin-stabilized satellite that revolves once every 3 s about a spin axis that is orthogonal to the ecliptic. It has a complex orbit but is often near the L1 Lagrangian point. The WAVES experiment employs three dipole antennas, one on the spin axis ($z$) and two orthogonal dipoles in the spin plane ($x, y$). The observations discussed here were made by the RAD1 and RAD2 radio receivers. RAD1 typically makes measurements at 32 frequencies, selected from 256 possible frequencies, distributed linearly from 20-1040 kHz whereas RAD2 makes measurements at 256 channels distributed linearly from 1.075-13.825 MHz. The data were downloaded from the WIND/WAVES web site\footnote{http://www-lep.gsfc.nasa.gov/waves/waves.html}. The RAD1 data discussed here use the $x$ dipole originally composed of two 50~m antenna segments, although in August 1998, one segment is believed to have lost roughly 15~m in length (M. Kaiser, private communication). The RAD2 data discussed here use the $y$ dipole composed of two 7.5~m antenna segments. The data have been averaged to a nominal time resolution of 1 min. 

The WIND/WAVES data were flux calibrated using the known parameters of the dipole antennas and receivers, and daily background measurements. The calibration was applied to events that have been previously published as calibrated spectra (e.g., Leblanc et al. 2000) and the results agreed well.  The calibrated data were also cross-checked against a second method that references the measurements to the galactic background radiation, which dominates the receiver noise over most of the RAD1 and RAD2 frequency ranges (Dulk et al. (2001). The two methods are in agreement to better than 3~dB for frequencies in the range 0.1-10 MHz,  the frequencies that concern us here.  

As will be shown, the type-II-like radio event analyzed here is quite faint. The typical background levels are $\sim 100$ and $>1000$ SFU in RAD1 and RAD2, respectively. It is therefore important to subtract the background from the spectra prior to analysis. For the purposes of display and analysis, it is convenient to interpolate the calibrated, background-subtracted spectra from RAD1 and RAD2 onto a single grid that displays the logarithm of the calibrated flux density as a function of time (linear abscissa) and the frequency (logarithmic ordinate). 

The observations are summarized in Fig.~2. The lower panel shows the WIND/WAVES dynamic spectrum over a frequency range of 100 kHz to 13.825 MHz and a time range of 500 min, starting at 21:40 UT on 2003 June 17. The log-flux is clipped at 1000 SFU to better show the radio emission of interest in the presence of other intense emissions. The upper panel shows the GOES 1-8\AA\ SXR light curve to illustrate the relative timing between the radio emissions and the associated SXR flare. The CME height-time measurements from SOHO/LASCO (Brueckner et al. 1995) were obtained from the NASA CDAW Data Center online CME catalog\footnote{http://cdaw.gsfc.nasa.gov/CME\_list (Yashiro et al. 2002)}. The CME shows little sign of acceleration or deceleration to an apparent height of nearly 30 R$_\odot$; a linear fit to the data (dotted line) yields an apparent speed of 1820 km s$^{-1}$. The mass and kinetic energy of the CME, also drawn from the online catalog, are $1.7\times 10^{16}$ gm and $2.8\times 10^{32}$ ergs, respectively, although these estimates are rather uncertain. Since the CME height observations are projected onto the sky plane, the measured CME speed represents a projected speed. Assuming the CME propagates radially from the position of the associated flare, and can be characterized by a hemispherical front and an opening angle $2\alpha$, an approximate correction factor can be derived (Leblanc et al. 2001; see also Sheeley et al. 1999). If the projected speed is $v_{sky}$ it should be multiplied by $(1+\sin\alpha)/(\sin\phi+\sin\alpha)$ to produce an estimate of the de-projected CME speed $v_{cme}$. The parameter $\phi$ is the angular distance of the associated active region from Sun center and $\alpha$ is the half-width of the opening angle of the CME. Inspection of Fig.~1 suggests $\alpha\approx 60^\circ$ is appropriate, yielding a deprojected speed $v_{cme}=2010$ km s$^{-1}$, although the correction factor is not very sensitive to $\alpha$. St.~Cyr et al. (2000) observed a mean value of $\alpha=36^\circ$, which would yield an estimated speed of 2050 km s$^{-1}$ in the present case.

The dynamic spectrum shows intense, fast-drift IP type III radio bursts during the rise phase of the SXR flare, with a peak flux density of $\sim 10^6$ SFU (1 SFU = $10^{-22}$ W m$^{-2}$ Hz$^{-1}$). Beginning at a time nearly coincident with the SXR maximum (22:55 UT), a faint, slow-drift radio event is seen, first detected at a frequency of $\approx 7.5$ MHz. The diamond symbol in the upper panel of Fig.~2 represents the projected radius of the CME ($\approx\!3.5$ R$_\odot$) at the estimated time of the onset of the radio event. During the course of more than 3 hrs, the emission drifts to lower frequencies and eventually becomes lost in the confusion of brighter emissions at a frequency of $\approx 300$ kHz at 01:30-02:30 UT on 2003 June 18, when the CME was $\sim\!30$ R$_\odot$ from the Sun.  Compared with the IP type III bursts, the IP type II-S event is very faint, with a peak flux density of only $\sim450$ SFU. The emission varies smoothly and shows neither substructure nor any sign of additional components that are harmonically related to the dominant lane of emission.  The variation of peak flux with time is characterized by a rapid rise to a broad maximum at $\approx\!23\!:\!10$ UT, followed by a monotonic decline to just a few SFU. If $\nu_{pk}(t)$ is the frequency of the  flux maximum at a time $t$ and $\Delta\nu(t)$ is the  FWHM bandwidth of the spectrum at that time, the bandwidth frequency ratio is $\Delta\nu/\nu_{pk}\approx 0.7\pm 0.1$. For reasons that will become apparent below, we refer to the slow-drift radio burst as an {\sl IP type II-S} event.

It is interesting to note the presence of a second type of emission beginning at approximately 02:30 UT on 2003 June 18 although it may begin with a faint fragment as early as 01:30 UT (upper band). Here, two lanes of emission are present. They are harmonically related and vary irregularly with time.  The frequency bandwidth of each lane is $\Delta\nu/\nu\approx 0.15$. The two lanes drift slowly to lower frequencies with time.  The emission in this case is identified as an {\sl IP type II-P} event. 

Note that while Cane \& Erickson (2005) cite the June 17-18 event as an "IP type II" event they show only the first 60 minutes of the RAD~2 dynamic spectrum. The RAD~1 spectrum shows type II-S emission extending down almost to 200 kHz, followed by the type II-P emission. Cane \& Erickson would have presumably classified the type II-P emission as "blobs and bands".

\section{Difficulties with the Plasma Radiation Hypothesis}

The emission identified here as an IP type II-S event differs in significant ways, both qualitatively and quantitatively, from most coronal type II bursts and their interplanetary extensions and analogs. While it matches the definition of an ``IP type II event", as defined by Cane \& Erickson (2005), the classification of IP type II-like emissions requires further refinement. In the case of the type II-S event, it is difficult to reconcile the properties of the radio emission with the assumption that plasma radiation is the relevant emission mechanism.

The frequency bandwidth ratio of the type II-S event is significantly larger than that typically observed for coronal and IP type II emissions ($\Delta\nu/\nu\sim 0.2-0.3$; e.g., Aguilar-Rodriguez et al. 2005), but is consistent with that of  ``IP type II events" although Cane \& Erickson did not analyze their sample quantitatively. Lengyel-Frey \& Stone (1989) noted the large bandwidths of certain IP type II radio emission (denoted ``class B" events by the authors) observed by the radio experiment on the ISEE-3 spacecraft. Aguilar-Rodriguez et al. argue that the large bandwidths inferred for ISEE-3 type II events may reflect a selection bias resulting from the frequencies sampled by the ISEE-3 experiment but nevertheless find (rare) events with frequency bandwidth ratios ranging from 0.5-0.8. 

Lengyel-Frey et al. (1989) suggested that large bandwidths can be understood in terms of density inhomogeneities in the source. If the density inhomogeneities are given by $\Delta n_e/n_e$, the bandwidth of the resulting plasma radiation should be $\Delta\nu/\nu = \Delta n_e/2n_e$. A problem with this idea is that the magnitude of plasma density fluctuations $\Delta n_e/n_e$ is typically quite small in the solar wind. Woo et al. (1995) used dual-frequency ranging from {\sl Ulysses} to show that $\Delta n_e/n_e$ varied from 1\% to no more than 20\% over periods of 20 min to 5 hrs (see also Celnikier 1987).  To account for the observed frequency bandwidth ratio on June 17-18 would require $\Delta n_e/n_e \sim 1.4$, much larger than is typically observed. It is therefore hard to see how a localized region could instantaneously produce broadband plasma radiation through random density inhomogeneities. If the source is very large the shock could encounter many discrete densities at any given time; but then it is difficult to understand why the distribution of emission is continous and smooth over the entire radial range that the event is observed.

More recently, Knock \& Cairns (2003, 2005) have quantitatively explored sources of spectral structure in coronal and IP type II bursts in the context of the plasma radiation model.  Knock \& Cairns (2005) consider the case of a shock expanding laterally in the quiescent corona and show that broadband emission can be produced, the two harmonics even merging in some cases. It is questionable whether such a model applies to the event considered here, however. By its very nature, a lateral shock in the corona will not propagate significantly in the radial direction and therefore does not propagate significantly into the interplanetary medium. Knock \& Cairns point out that the frequency drift rate resulting from lateral shock expansion would be much less than that resulting from a radially propagating shock. As shown below, the type II-S event discussed here shows a frequency drift that is significantly {\sl faster} than can be accounted for by plasma radiation driven by a radially propagating shock.  

Indeed, a defining property of coronal and IP type II radio emission is the drift of the characteristic frequency from higher to lower values with time. The frequency and drift rate of type II radio emission are easily measured and, in the context of plasma radiation, can be interpreted in a straightforward manner. The measured frequency is assumed to be a measure of the electron plasma frequency $\nu_{pe}$ and the drift rate is therefore assumed to be $\dot{\nu}\propto v_S \nabla n_e/\nu_{pe}$ where $v_S$ is the speed of the source parallel to the density gradient. The most common interpretation of IP type II radio emission is that it occurs in the foreshock region of the fast CME (see Cairns et al. 2000 for a summary of the data and arguments in support of this conclusion). The density in the source is therefore assumed to be the relatively undisturbed corona and/or interplanetary medium. Analysis of coronal and IP type II radio emissions typically involves fitting the time evolution of the spectrum to a shock speed and trajectory, and a density model. Semi-empirical models based on white light (e.g.,  Newkirk 1967; Saito 1970; Saito et al. 1977) or radio data (Fainberg \& Stone 1971; Bird et al. 1994; Leblanc et al. 1998) are employed, although the model is often renormalized by a constant scaling factor (e.g., Reiner et al. 2003). Adopting a given density model, the source speed can then be inferred (e.g., Kaiser et al. 1998) from the frequency drift of the type II emission. With the availability of high quality white light coronagrams from SOHO/LASCO over wide range of coronal heights in recent years, the projected speed of the shock driver -- the CME -- is known. Recent work has used radio and white light observations jointly to constrain shock dynamics (Reiner et al. 2003). Reiner et al. (1998) point out that the density varies with radius nearly as $r^{-2}$ beyond a few solar radii, so the plasma frequency and hence, the observed radio frequency, should vary as $r^{-1}$, an expectation that is often borne out.

In the present case, the projected speed of the CME is well-measured; the start time and start frequency of the type II-S event are also well-constrained. Hence the initial radius $r_\circ$, corrected for projection, and the electron number density $n_e(r_\circ)$ are presumed known if the source is associated with the CME shock and plasma radiation is the relevant emission mechanism. To be concrete, the density is assumed to vary with radius according to the model of Saito et al. (1977) although other density models yield similar results. Starting at $t_\circ$ the CME is assumed to propagate from $r_\circ$ radially outward with either a projected speed of 1820 km s$^{-1}$ (case 1) or a de-projected speed of 2010 km s$^{-1}$ (case 2), driving a shock which produces plasma radiation. Fig.~3a shows the expected drift rate for the Saito density model overlaid on the dynamic spectrum for case 1 (dash-dot) and case 2 (dashed). In order to match the initial condition, the Saito model must be multiplied by a factors of $\approx\!4$ and 10 for cases 1 and 2, respectively, assuming fundamental plasma radiation. These normalization values should be divided by 4 if harmonic plasma radiation is assumed. Regardless, the time variation of plasma radiation fails to match that of the type II-S event. Indeed, in order to approximately match the frequency drift, the electron density must vary as $n_e\propto r^{-3}$ (case 1) or $r^{-2.75}$ (case 2), and to account for the initial condition the source must be overdense relative to Saito at $r_\circ$ by factors of 13 and 6.5, respectively (fundamental emission). To account for the type II-S event in terms of a plasma radiation model therefore requires rather unusual conditions in the inner heliosphere: the electron number density must be overdense relative to the Saito et al (1997) model, yet decline with radius as $\sim r^{-3}$ out to $\gtrsim 30$ R$_\odot$ rather than rapidly tending toward the expected $r^{-2}$ dependence. 

In contrast to the type II-S event, the emission identified as type II-P in Fig.~2 and described in \S2, is entirely consistent with plasma radiation: it shows both fundamental and harmonic bands, the bands have narrow frequency bandwidth ratios, the intensity varies irregularly with time, and the emission drifts slowly to lower frequencies with time. The flux density of the fundamental band is $\sim 150$ SFU while the harmonic emission is considerably weaker, a property that has been noted previously for IP type II harmonic pairs (Lengyel-Frey et al. 1985; Lengyel-Frey \& Stone 1989). The frequency drift of the type II-P emission is also consistent with plasma radiation: it can be described by plasma radiation from a Saito density model, with no renormalization needed, if the shock launches at the same time as the CME, but propagates at $\approx 1200$ km s$^{-1}$ . This might occur if the shock front responsible for the type II-P emission propagates at an angle $50-60^\circ$ from the radial, as would be the case if the source is near the flank of the CME rather than the nose. 

To summarize, the fast halo CME of 2003 June 17 produced a type-II-like radio event, referred to here as an IP type II-S event, that is characterized by a smoothly varying, relatively broadband, single lane of emission. The large bandwidth and the smooth variation of the radio emission with time are difficult to understand in the context of plasma radiation. If plasma radiation is nevertheless the relevant emission mechanism, then rather special conditions in the IPM would be required to account for the variation of radio frequency with time; namely, a density model that is initially overdense relative to the model of Saito et al. (1977), declines with radius significantly more steeply than a Saito model out to at least 30 R$_\odot$, and must contain extreme and rather uniform density variations over the full radial range to account for the large bandwidth.  In contrast, the emission referred to here as IP type II-P is entirely consistent with plasma radiation from a shock propagating through a Saito et al density model, but only if the shock is associated with a shock near the flank of the fast CME. 

\section{Synchrotron Radiation}

In this section an alternative to plasma radiation from an IP shock driven by a fast CME is considered: incoherent synchrotron radiation. Synchrotron radiation is emitted by energetic electrons gyrating in an ambient magnetic field. It is a well-understood mechanism that is believed to play a central role in a wide variety of astrophysical phenomena, including solar flares (e.g., Bastian et al. 1998), supernova remnants (e.g., Reynolds \& Chevalier 1981), and extragalactic radio sources (e.g., Begelman et al. 1984). The observed spectrum of sychrotron radiation depends on the magnetic field, the electron distribution function, the ambient plasma in the source, and the medium external to the source. The frequency bandwidth ratio of a self-absorbed synchrotron emission spectrum is $\Delta\nu/\nu\sim 1$ but it can be smaller. The low-frequency spectrum can cutoff steeply if Razin suppression and/or thermal free-free absorption are operative; the high frequency spectrum can cutoff if the electron energy distribution has a high energy cutoff (see, e.g., Ramaty 1969, Ramaty \& Petrosian 1972, Klein 1987). 

As noted previously, the frequency drift with time from high to low values is a defining characteristic of type II radio emission. Can synchrotron emission produce a frequency change in time similar to that observed? Consider a power-law distribution of electrons with a number density $n_{rl}(E)dE\propto E^{-\delta}dE$. The spectral maximum of the emission spectrum then occurs at a frequency $\nu_{pk}\propto n_{rl}^{2/(\delta+4)}B^{(\delta+2)/(\delta+4)}$ (Dulk 1985), where $B$ is the magnetic field. It is  clear that if $n_{rl}$ and/or $B$ decrease in the source with time, then so does $\nu_{pk}$. For example, if $n_{rl}$ and $B$ both vary as  $\sim r^{-2}$, then so does $\nu_{pk}$. I present a more quantitative comparison below. 

Another question: is the flux density of the source commensurate with the proposed emission mechanism? A limit to the brightness temperature $T_B$ of a self-absorbed synchrotron source is imposed by inverse Compton scattering, constraining it to be no more than $10^{11}-10^{12}$ K (Kellermann \& Pauliny-Toth 1969). Observations of certain coronal type II radio bursts yield brightness temperatures well in excess of this limit (Nelson \& Melrose and references therein), thereby eliminating synchrotron radiation as the relevant emission mechanisms on these grounds alone. The flux density of the type II-S source is related to its brightness temperature by $S=2 k_B T_B \nu^2\Omega/c^2$ where $k_B$ is Boltzmann's constant and $\Omega$ is the solid angle subtended by the source. While WIND/WAVES measured the flux density of the type II-S event, no direct information is available on its angular size. However, if it is assumed that the source is comparable in size to the projected size of the CME itself, a maximum brightness temperature can be inferred and compared to the inverse Compton limit. The IP type II-S event had a maximum flux density of $\approx 450$ SFU at about 23:10 UT, when the projected height of the CME was $r\approx 6$ R$_\odot$ and the frequency is $\approx 3$ MHz. Taking the solid angle subtended by the source to be $\Omega\sim r^2/D^2$, where $D=1$ AU, $T_B\sim 10^{10}$ K, well below the Compton limit. Therefore, the possibility that synchrotron emission is relevant to type II-S events is not excluded on the basis of the observed flux and the inferred brightness temperature.

To pursue the idea further, a simple model was developed as a means of illustrating that type II-S bursts can be ascribed to synchrotron radiation from near-relativistic electrons. It is not intended to explain the familiar type II-P emissions for which the plasma radiation mechanism is assumed to be relevant. I assume that, just as is the case for IP type II-P emission, type II-S radio events are causally related to fast halo CMEs. In other words, I assume that the type II-S source region is closely associated with the fast CME. As the CME and the associated type II-S source increase their distance from the Sun, the source size, magnetic field, plasma density, and number of energetic electrons are all assumed to vary. For illustrative puposes, the ingredients of the schematic model refer to the event of 2003 June 17 and are as follows:

\begin{enumerate}
\item {\sl Source speed}: The source is assumed to move radially outward from the Sun at a constant speed $v_{cme}$.
\item {\sl Source size}: The type II-S source size $s$ is assumed to increase linearly with time $t$ so that the solid angle subtended by the source increases approximately as $t^2$. Specifically, I take $s\sim r=r_\circ+v_{cme}t$, where $r_\circ$ is the initial source size at $t=0$ when the type II-S event begins. The angular size of the source is computed as $\Omega=A/(D-r)^2$, where $A=\pi s^2$ is the source area and $D$ is 1 AU. The source depth is taken to be $L=0.1r$ (e.g., Manchester et al. 2005). 
\item {\sl Aspect angle}: Implicit in the expression for the source size is the assumption that the source is viewed approximately ``head-on", as appropriate to a halo CME (although the bulk of the June 17-18 event in fact propagated to the east). It is further assumed that the magnetic field is oriented $60^\circ$ to the line of sight for the purposes of this illustrative calculation.  
\item {\sl Plasma density}: The unperturbed solar wind plasma density is assumed to vary  according to a Saito et al. (1977) model. It is found that the best results were obtained when the source volume was {\sl underdense} relative to Saito et al. (see below). 
\item {\sl Magnetic field}: The magnetic field in the source is assumed to vary as $B=B_\circ (r/R_\odot)^{-\beta}$.
\item {\sl Energetic electrons}: The electron distribution is assumed to be power-law in energy, isotropic in pitch-angle, and to fill the source uniformly: $n_e(r,E)dE=K(r) E^{-\delta} dE$, where $K(r)$ embodies the radial variation in the number density of energetic electrons. Furthermore, the distribution is characterized by a low-energy cutoff of $E_1$, to which the emission spectrum is insensitive, and a high energy cutoff $E_2$, to which the emission spectrum is somewhat sensitive. Therefore, the total number of energetic electrons $n_{rl}$ between $E_1$ and $E_2$ is $n_{rl}(r)=K(r)\int_{E_1}^{E_2} E^{-\delta} dE$. I assume $E_1=100$ keV, $E_2=1$ MeV, and $\delta=5$. Finally, it is assumed that $K(r)$ increases linearly with time from 0.001\% to a few times 0.1\% of the ambient plasma density until roughly 02:00 UT.
\end{enumerate}

\noindent A computer program has been written that embodies these assumptions, a modified version of the code described by Bastian et al. (2001). Briefly, the source radius is computed once each minute according to the assumed speed of the CME and the corresponding values of the magnetic field, ambient plasma density, energetic electron density, source area, and source depth are derived. The source flux density is then computed at a given frequency using 

\begin{equation}
S(\nu, B, \theta, T, n_{th}, E_1, E_2, n_{rl}, \delta)= 2 k_B
{\nu^2\over c^2} {1\over{\mu^2(\nu,n_{th},B,\theta)}}{j_T \over
\kappa_T}{A\over (D-r)^2}[1-\exp{(\kappa_TL)}]
\end{equation}

\noindent where $\mu(\nu,n_{th},B,\theta)$ is the index of refraction, $j_T = j_{ff}(\nu, T, n_{th})+j_{s}(\nu,B, \theta,n_{th},n_{rl},E_o, E_c,\delta)$ is the total emissivity due to the free-free and synchrotron mechanisms, $\kappa_T = \kappa_{ff}(\nu,T,n_{th})+\kappa_{s}(\nu,B,\theta,n_{rl},E_o,E_c,\delta)$ is the total absorption coefficient, $A/(D-r)^2$ is the solid angle subtended by the source, and $D$ is the distance from the observer to the source (1 AU). In the case considered here thermal free-free absorption is negligible for the parameters chosen. The expressions for  $j_s$ and $\kappa_s$ for the case of mildly relativistic electrons are quite cumbersome (Ramaty 1969, Benka \& Holman 1992) and their evaluation is computationally demanding. They were instead calculated using 
using the approximate expressions of Klein (1987), which are accurate to better than 10\% over the range of conditions considered. Model spectra were computed using calculaions at 100 frequencies distributed logarithmically between 0.1-13.825 MHz and displayed as a dynamic spectrum identical in format to that shown in Fig.~1. The results are shown in Figs.~3b and 3c. Two cases were considered. In the first case, the source was assumed to propagate out from the Sun with a speed $v_{cme}=1820$ km s$^{-1}$, the apparent speed on the sky (case 1). A good fit to the data is obtained when $B_\circ=4$ G and $\beta=1.75$. However, it was necessary to assume that the ambient density was underdense relative to a Saito model by a factor of 4. In the second case, it was assumed that the source propagated out from the Sun with a speed $v_{cme}=2010$ km s$^{-1}$, the deprojected speed (case 2). In this case, a good fit to the data was obtained when $B_\circ=3$ G and $\beta=1.65$, and an ambient density that is underdense relative to a Saito model by a factor of 3. The reason the ambient density must be underdense relative to a Saito model is that Razin suppression would otherwise cut off the emission at a higher frequency than is observed. The run of magnetic field in the source from 3-30 R$_\odot$ is roughly 0.5 to 0.01 G in both cases. It is interesting to note that Bastian et al. (2001) fit a magnetic field strength of 0.33 G at a radius of 2.8 R$_\odot$ in the case of the fast CME observed on 1998 April 20. 

Fig.~3b shows the model dynamic spectrum for case 1; the case 2 spectrum is nearly indistinguishable from case 1 visually. Fig.~3c again reproduces the WIND/WAVES dynamic spectrum. This time, the plots overlaid on the type II-S event show the variation of the peak frequency $\nu_{pk}$ of the model synchrotron spectrum as a function of time. They follow the observed variation in frequency with time rather well in both cases. However, the peak flux density of the models falls short of that measured by a factor of a few. A straightforward modification would be to simply increase the diameter of the source. Alternatively, or in addition, the details of the electron distribution function, the time variation of the number of energetic electrons, and the magnetic field could be ``fine tuned", an exercise I do not pursue here. Additional examples of the IP type II-S phenomenon will be analyzed and modeled in a forthcoming publication. 

It is important to emphasize that type II-S events potentially offer important new diagnostics of physical conditions in the source, leveraging information about the magnetic field, the plasma density, and the distribution of energetic electrons in the source. With the launch of STEREO/WAVES, important new observations of these sources are possible such as more precise constraints on the location and size of the source. The stereoscopic observations provided by STEREO may also provide constraints on the directionality of the emission, which can be significant for plasma radiation, but is not expected to be a factor for synchrotron radiation as long as the electron distribution function is nearly isotropic.

\section{Discussion and Conclusions}

Observations of an IP type-II-like radio event that occurred on 2003 June 17-18 in association with a fast halo CME, a strong SXR flare, and a solar proton event have been presented. It is difficult to account for the observed properties of the event in terms of plasma radiation. Indeed, the properties of this IP type II-S event can be adequately accounted for in terms of synchrotron emission from a source associated with the fast CME.  In contrast, the emission referred to here as type II-P remains fully consistent with plasma radiation from the shock if it is associated with the flank of the CME. Therefore, it may be necessary to distinguish between slow-drift radio bursts that can be attributed to plasma radiation (IP type II-P) and those that can be attributed to synchrotron radiation (IP type II-S). Since the IP type II-S event shares many attributes with Cane \& Erickson's ``IP type II events", and ``IP type II events" are relatively rare, IP type II-S events are presumably rarer still, since it remains to be seen what fraction of Cane \& Erickson's ``IP type II events" can be attributed to synchrotron radiation.  Future attempts to classify IP type II or type-II-like phenomena will need to distinguish between the underlying radiation mechanisms.

It is worth asking, at this point, whether the designation "type II-S" is the best descriptor of the event analyzed here. Is it possible that the observations are better described as the interplanetary analog to a "moving type IV burst", or type IVm? Type IVm radio bursts are a rare coronal phenomenon that, despite decades of observations, remain poorly understood (Stewart 1985 and references therein). They have been attributed to both plasma radiation and/or synchrotron radiation. They have been classified (Smerd \& Dulk 1971) as isolated sources, expanding arches, and advancing fronts (Kai 1970). The range of speeds inferred for type IVm sources is quite similar to that of CMEs, ranging from 200-1600 km s$^{-1}$. The "advancing front" type IVm has been attributed to a shock that produces a coronal type II radio followed synchrotron radiation beyond 2R$_\odot$. It may be that the "advancing front" type IVm is analogous to the June 17-18 event analyzed here.  Only one example of type IVm has been reported in the WIND RAD1 and RAD2 band (Reiner et al. 2006), a broadband, highly circularly polarized source that showed a slow frequency drift from 2 MHz to 0.8 MHz over a period of four days. The spectrum is strikingly different from the present case. Given the persistent uncertainties in interpreting type IVm bursts and the fact that both Kaiser's online catalog and Cane \& Erickson classify the event of 2003 June 17-18 as a "type II", I retain the terminology type II-S as a meaningful descriptor. While it now appears that IP type-II-like emissions may involve two distinct emission mechanisms, both the type II-P and type II-S emission remain closely associated with fast CMEs.

In view of the likely role of synchrotron radiation in type II-S emission, a number of more fundamental questions are raised by the event on 2003 June 17-18: 

\begin{itemize}
\item Where is the source region of the type II-S relative to the fast CME?
\item What is the source of the energetic electrons that emit the synchrotron radiation and how are they transported throughout the source?
\item What are the relevant electron acceleration and loss mechanisms?
\end{itemize}

\noindent While these questions are left open here, I nevertheless comment briefly on each. Regarding the location of the type II-S relative to the CME, it is widely assumed that the source region of type II-P events is in the foreshock region of a shock driven by an interplanetary CME (Cairns et al. 2000). I consider the foreshock region as an unlikely source region for type II-S events for the simple reason that the undisturbed solar wind magnetic field in the foreshock region will be significantly weaker than in the postshock region (Manchester et al. 2005). Since the synchrotron emissivity $j_s$ depends on the magnetic field roughly as $j_s \sim B^{5/2}$ a given population of energetic electrons will emit far more strongly in the postshock region than in the foreshock region if they are present there. Therefore, one possible source region is the sheath between the CME and the shock it drives. The fact that the best model fit resulted from an ambient density that is underdense relative to Saito et al. (1997) by a factor of 3 may present difficulties in this case because both the density and the magnetic field are expected to be enhanced in the sheath. In addition, the electrons responsible for the synchrotron emission may escape rather easily from the sheath region.

A second possibility is the CME flux rope, or magnetic cloud (Burlaga et al 1981).  It is interesting to compare the inferred magnetic field variation in the source with that inferred on observational and theoretical grounds. A study of a large sample of interplanetary CMEs (ICMEs) by Wang et al. (2005) suggests that $\beta\approx 1.5$ On the other hand, for self-similar flux rope models (Gibson \& Low 1998), $\beta=2$. These values neatly bracket those inferred from the synchrotron model. If the source region is associated with the CME flux rope, it might be regarded as analogous to the fast CME on 1998 April 20 described by Bastian et al (2001), which displayed synchrotron radiation from MeV electrons entrained in the CME magnetic loops, albeit at much higher frequencies (164-432 MHz) than those considered here, observed to a height of $\approx 3.5$ R$_\odot$.  If the source electrons reside in the flux rope, the fact that the source is underdense may not be a problem. 

While the source of the energetic electrons responsible for the synchrotron emission is unknown, as are the acceleration and transport processes, the fact that such electrons occur in association with fast CMEs and large solar flares is well established (see the recent review by Kahler 2007, and references therein).  The energy requirements of the type II-S event in the context of synchrotron radiation from near-relativistic electrons does not appear to present difficulties from an observational perspective. Mewaldt et al. (2005) analyzed the energy contained in energetic protons, helium, and electrons associated with the 2003 Oct-Nov events and conclude that accelerated IP particles amount to 0.4\% to as much as 24\% of the energy of the associated CME and that from 1\% to 18\% of the particle energy is contained in electrons. One possibility for the source of the electrons is the high Mach number shock driven by the CME (Mann et al. 1999, 2002). If the source of the type II-S event is in the sheath region, then shock-accelerated electrons have ready access and may therefore be relevant. However, if the source region is in the flux rope, it is less clear that shock acceleration is relevant because it is not known how shock-accelerated electrons could then gain access to the flux rope. Alternatives to the CME-driven shock include electron acceleration in the flare (e.g., Simnett et al. 2005, 2006), in the current sheet following the CME (e.g., Cliver et al. 1986), and/or sustained electron acceleration and release in the solar corona (Klein et al. 2001, 2005). The presumption in these cases is that energetic electrons are fed into the expanding ICME for an extended period of time, replenishing those electrons that lose their energy to adiabatic losses (synchrotron cooling is unimportant). The question of whether these are viable alternatives or not is beyond the scope of this paper. 

To conclude, the IP type II-S event may represent a new class of interplanetary radio phenomenon. Just as is the case for conventional IP type II-P events, the type II-S event is closely associated with a fast CME. However, unlike IP type II-P events, the type II-S is interpreted as synchrotron emission from relativistic electrons interacting with a magnetic field. Such electrons are far more energetic than those responsible for plasma radiation. The precise location of the type II-S emission is presently unknown. I have suggested the post-shock sheath region or the flux rope as possible source locations. The source of the relativistic electrons is also unknown; I have suggested the fast CME-driven shock, the flare, CME current sheet, or other energy release processes in the solar corona as possible sources. To refine the nature of IP type II-S events, to understand their relation to other coronal and IP bursts, and to answer fundamental questions raised above, requires the identification and analysis of many more events. A more comprehensive study of additional events observed by WIND/WAVES will be forthcoming. The recently launched STEREO/WAVES experiment should provide important new data, including constraints on the source size and direction. 

\acknowledgements{I thank M. Kaiser for discussion of WIND/WAVES RAD1 and RAD2 calibration issues and for maintaining the web site from which these data were obtained. I thank G. Dulk for discussion and S. Kahler for comments on a preliminary draft of this paper. The online CME catalog is generated and maintained at the CDAW Data Center by NASA and The Catholic University of America in cooperation with the Naval Research Laboratory. SOHO is a project of international cooperation between ESA and NASA. }

\begin{figure}
\epsscale{0.8}
\plotone{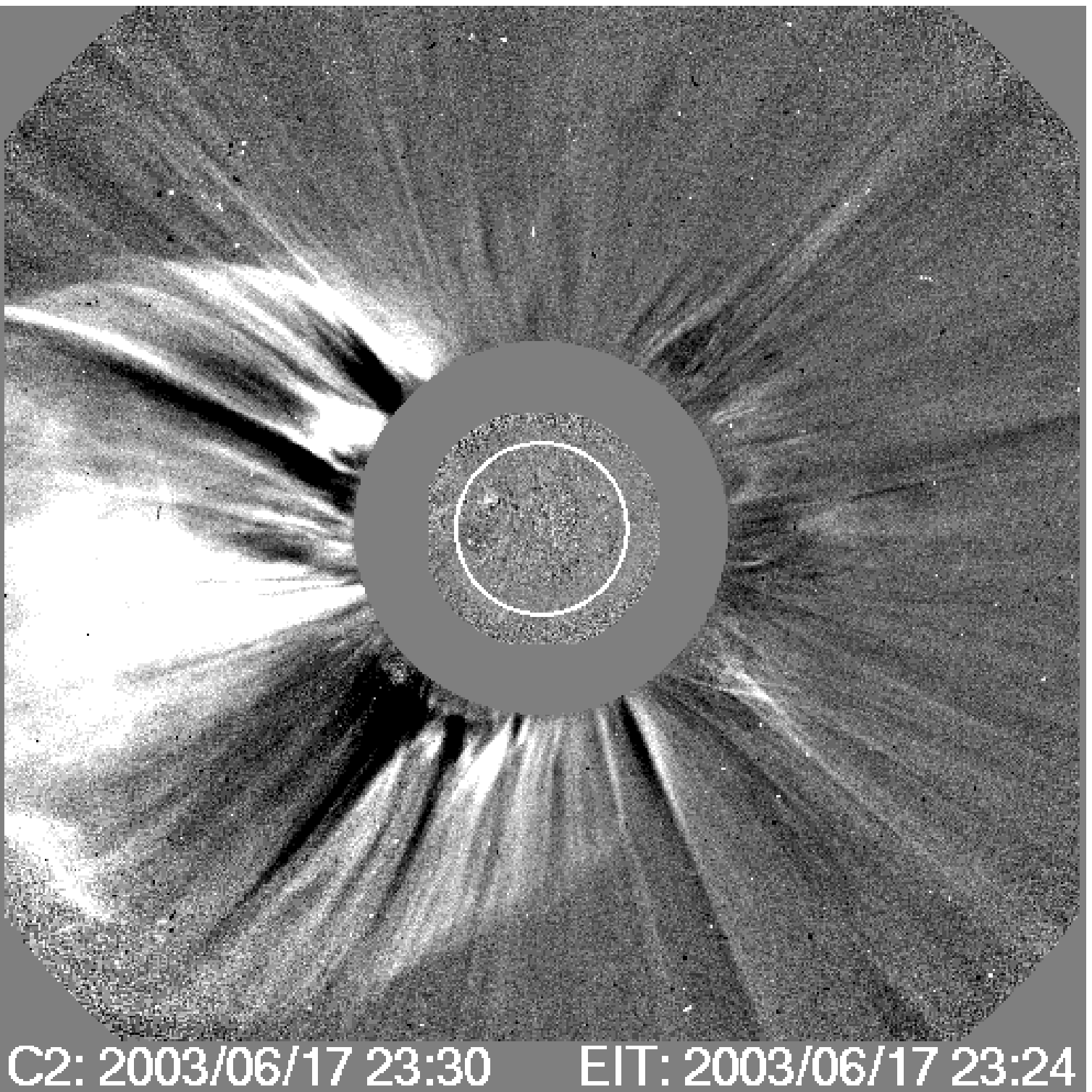}
\caption{The fast halo CME of 2003 June 17 is shown as a SOHO/LASCO C2 difference image at 23:30:05 UT.}
\end{figure}

\begin{figure}
\plotone{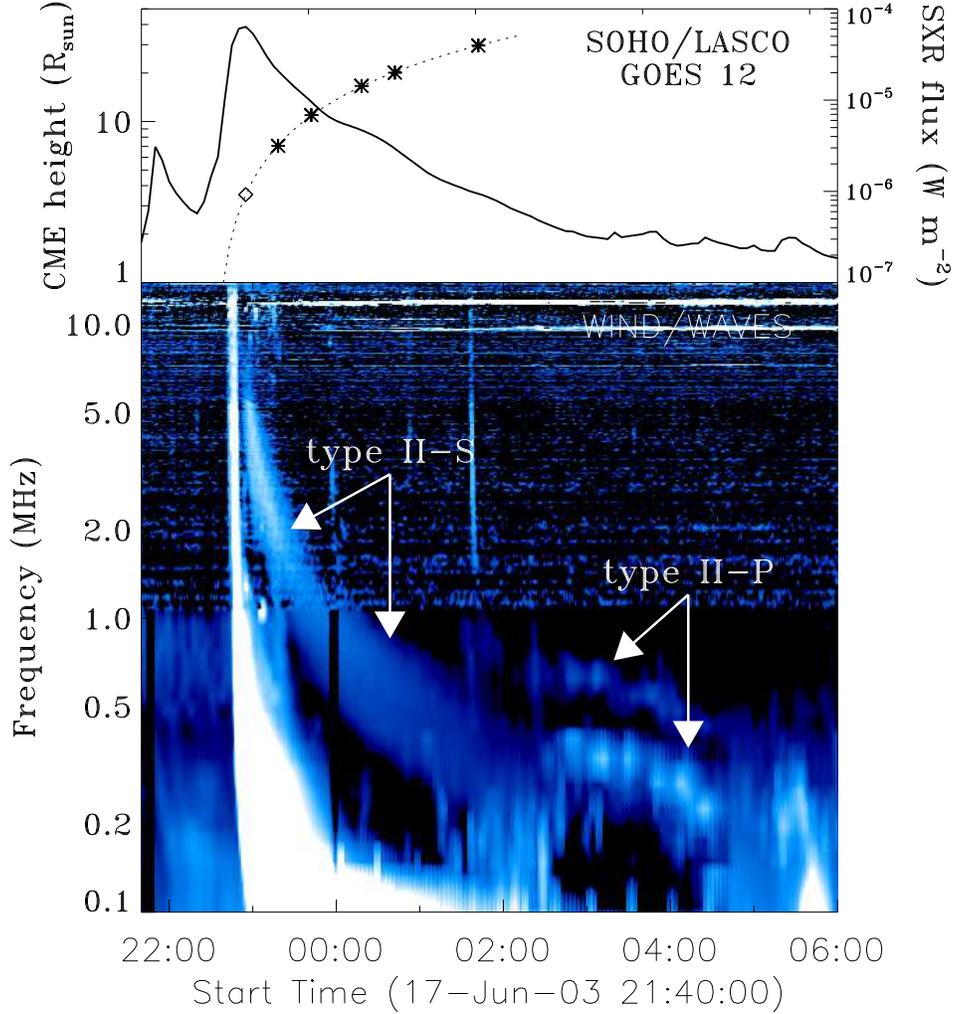}
\caption{Observational summary of radio and SXR emission from the fast halo CME on 2003 June 17. The top panel shows the time variation of the GOES 1-8\AA\  SXR flux (right-hand axis) and the apparent height of the associated CME as a function of time (asterisks), as measured by SOHO/LASCO (left-hand axis). The dashed line represents a linear fit to the CME data. The diamond symbol shows the inferred height of the CME at the time the type II-S event begins. The lower panel shows a calibrated, background-subtracted dynamic spectrum composed from WIND/WAVES RAD1 and RAD2 data. The type II-S event appears as a diffuse, slow-drift band following the type III radio bursts. The type II-P emission appears as a harmonic pair of narrow-band lanes following the type II-S event.}
\end{figure}

\begin{figure}
\epsscale{0.4}
\plotone{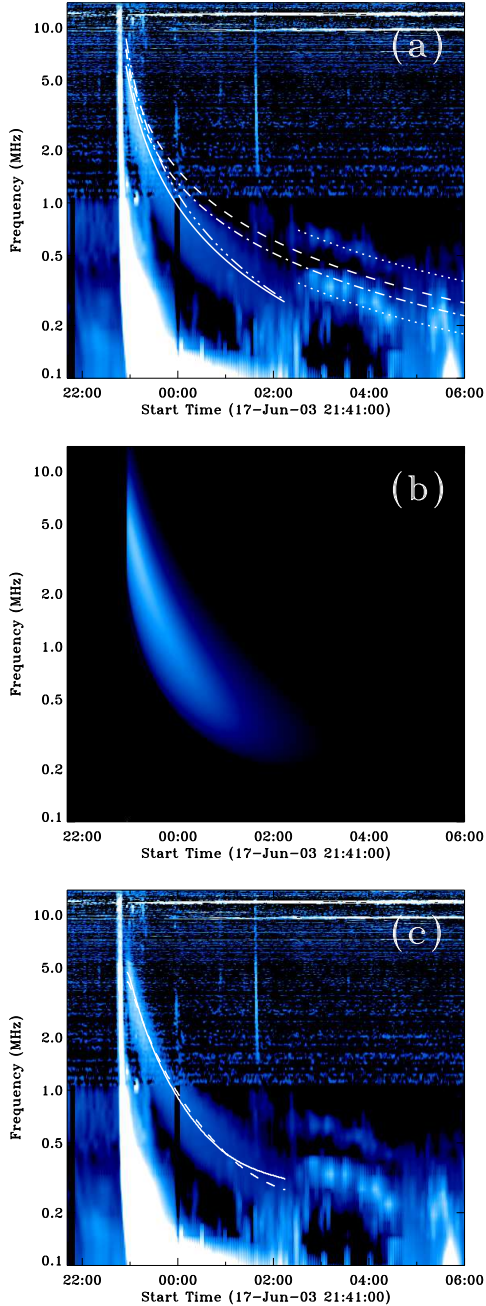}
\caption{a) Comparison of the frequency drift rates expected from plasma radiation from a Saito et al. (1977) density model. The type II-S emission cannot simultaneously fit the start frequency and the frequency drift (dashed, dot-dash lines). However, the type II-P fundamental and harmonic emission is adequately fit (dotted lines) by the Saito et al density model if the shock is on the flank of the CME. The type II-S can be approximately fit by a density models that varies as $r^{-3}$ (dots-dash line) or $r^{-2.75}$ (solid line). b) A simple synchrotron model of the type II-S event. See text for a discussion of the model assumptions and parameters. c) same as panel (a), but with plots of $\nu_{pk}(t)$ resulting from the synchrotron models. The solid line uses the deprojected CME speed (2010 km s$^{-1}$) while the dashed line uses the projected speed (1820 km s$^{-1}$).  }
\end{figure}


\begin{thebibliography}{}

\bibitem[Aguilar-Rodriguez et al.(2005)]{2005JGRA..11012S08A} 
Aguilar-Rodriguez, E., Gopalswamy, N., MacDowall, R., Yashiro, S., \& 
Kaiser, M.~L.\ 2005 \jgr\ 110, 12 

\bibitem[Aguilar-Rodriguez et al.(2005)]{2005soho...16..393A} 
Aguilar-Rodriguez, E., Gopalswamy, N., MacDowall, R., \& et al.\ 2005, ESA 
SP-592: Solar Wind 11/SOHO 16, Connecting Sun and Heliosphere, 16, 393 


\bibitem[Bastian et al.(1998)]{1998ARA&A..36..131B} Bastian, T.~S., Benz, 
A.~O., \& Gary, D.~E.\ 1998, \araa, 36, 131 

\bibitem[Bastian et al.(2001)]{2001ApJ...558L..65B} Bastian, T.~S., Pick, 
M., Kerdraon, A., Maia, D., \& Vourlidas, A.\ 2001, \apjl, 558, L65 

\bibitem[Begelman et al.(1984)]{1984RvMP...56..255B} Begelman, M.~C., 
Blandford, R.~D., \& Rees, M.~J.\ 1984, Reviews of Modern Physics, 56, 255 

\bibitem[Bird et al.(1994)]{1994ApJ...426..373B} Bird, M.~K., Volland, H., 
Paetzold, M., Edenhofer, P., Asmar, S.~W., \& Brenkle, J.~P.\ 1994, \apj, 
426, 373 

\bibitem[Boischot et al.(1980)]{1980SoPh...65..397B} Boischot, A., Riddle, A.~C., Pearce, J.~B., \& Warwick, J.~W.\ 1980, \solphys, 65, 397 

\bibitem[Bougeret et al.(1984)]{1984A&A...141...17B} Bougeret, J.-L., 
Fainberg, J., \& Stone, R.~G.\ 1984, \aap, 141, 17 

\bibitem[Bougeret et al.(1995)]{1995SSRv...71..231B} Bougeret, J.-L., et 
al.\ 1995, Space Science Reviews, 71, 231 

\bibitem[Burlaga et al. 1988]{Bur88} Burlaga, L. F., Sittler, E., Mariani, F., \& Schwen, R. 1981, \jgr 86, 6673

\bibitem[Brueckner et al.(1995)]{1995SoPh..162..357B} Brueckner, G.~E., et 
al.\ 1995, \solphys, 162, 357 

\bibitem[Cairns et al.(2000)]{2000PASA...17...22C} Cairns, I.~H., Robinson, 
P.~A., \& Zank, G.~P.\ 2000, Publications of the Astronomical Society of 
Australia, 17, 22 

\bibitem[Cane et al.(1982)]{1982SoPh...78..187C} Cane, H.~V., Stone, R.~G., 
Fainberg, J., Steinberg, J.~L., \& Hoang, S.\ 1982, \solphys, 78, 187 

\bibitem[Cane(1984)]{1984A&A...140..205C} Cane, H.~V.\ 1984, \aap, 140, 205 


\bibitem[Cane(1985)]{1985JGR....90..191C} Cane, H.~V.\ 1985, \jgr, 90, 191 


\bibitem[Cane \& Reames(1988)]{1988ApJ...325..895C} Cane, H.~V., \& Reames, 
D.~V.\ 1988, \apj, 325, 895 


\bibitem[Cane \& Erickson(2005)]{2005ApJ...623.1180C} Cane, H.~V., \& 
Erickson, W.~C.\ 2005, \apj, 623, 1180 

\bibitem[Celnikier et al.(1987)]{1987A&A...181..138C} Celnikier, L.~M., 
Muschietti, L., \& Goldman, M.~V.\ 1987, \aap, 181, 138 



\bibitem[Cliver et al.(1986)]{1986ApJ...305..920C} Cliver, E.~W., Dennis, 
B.~R., Kiplinger, A.~L., Kane, S.~R., Neidig, D.~F., Sheeley, N.~R., Jr., 
\& Koomen, M.~J.\ 1986, \apj, 305, 920 

\bibitem[Cliver et al.(1999)]{1999SoPh..187...89C} Cliver, E.~W., Webb, 
D.~F., \& Howard, R.~A.\ 1999, \solphys, 187, 89 

\bibitem[Cliver et al.(2004)]{2004SoPh..225..105C} Cliver, E.~W., Nitta, 
N.~V., Thompson, B.~J., \& Zhang, J.\ 2004, \solphys, 225, 105 

\bibitem[Dulk(1985)]{1985ARA&A..23..169D} Dulk, G.~A.\ 1985, \araa, 23, 169 

\bibitem[Dulk et al.(1999)]{1999GeoRL..26.2331D} Dulk, G.~A., Leblanc, Y., 
\& Bougeret, J.-L.\ 1999, \grl, 26, 2331 

\bibitem[Fainberg, J., \& Stone, R. 1971]{Fai78} Fainberg, J., \& Stone, R. G. 1971, Sol. Phys. 17, 392

\bibitem[Fainberg et al.(1985)]{1985A&A...153..145F} Fainberg, J., Hoang, 
S., \& Manning, R.\ 1985, \aap, 153, 145 

\bibitem[Gibson \& Low 1998]{Gib98} Gibson, S., \& Low, B. C. 1998, \apj 493, 460 


\bibitem[Gopalswamy et al.(1997)]{1997ApJ...486.1036G} Gopalswamy, N., 
Kundu, M.~R., Manoharan, P.~K., Raoult, A., Nitta, N., \& Zarka, P.\ 1997, 
\apj, 486, 1036 


 


\bibitem[Gopalswamy(2006)]{2006GMS...165..207G} Gopalswamy, N.\ 2006, 
Washington DC American Geophysical Union Geophysical Monograph Series, 165, 
207 




\bibitem[Kahler 2007]{Kah07} Kahler, S.~W. 2007, Sp. Sci. Rev. 

\bibitem[Kai(1970)]{1970SoPh...11..310K} Kai, K.\ 1970, \solphys, 11, 310 

\bibitem[Kaiser et al.(1998)]{1998GeoRL..25.2501K} Kaiser, M.~L., Reiner, 
M.~J., Gopalswamy, N., Howard, R.~A., St.~Cyr, O.~C., Thompson, B.~J., \& 
Bougeret, J.-L.\ 1998, \grl, 25, 2501 

\bibitem[Kellermann \& Pauliny-Toth(1969)]{1969ApJ...155L..71K} Kellermann, 
K.~I., \& Pauliny-Toth, I.~I.~K.\ 1969, \apjl, 155, L71 


\bibitem[Klein(1987)]{1987A&A...183..341K} Klein, K.-L.\ 1987, \aap, 183, 341

\bibitem[Klein et al.(2001)]{2001A&A...373.1073K} Klein, K.-L., Trottet, 
G., Lantos, P., \& Delaboudini{\`e}re, J.-P.\ 2001, \aap, 373, 1073 

\bibitem[Klein et al.(2005)]{2005A&A...431.1047K} Klein, K.-L., Krucker, 
S., Trottet, G., \& Hoang, S.\ 2005, \aap, 431, 1047 


\bibitem[Knock et al.(2003)]{2003JGRA.108j.SSH2K} Knock, S.~A., Cairns, 
I.~H., \& Robinson, P.~A.\ 2003, Journal of Geophysical Research (Space 
Physics), 108, 2 


\bibitem[Knock \& Cairns(2005)]{2005JGRA..11001101K} Knock, S.~A., \& 
Cairns, I.~H.\ 2005, Journal of Geophysical Research (Space Physics), 110, 
1101 


\bibitem[Leblanc, Y., Dulk, G. A., Bougeret, J. 1998]{Leb98} Leblanc, Y., Dulk, G. A., \& Bougeret, J.-L. 1998, Sol. Phys., 183, 165

\bibitem[Leblanc et al.(2001)]{2001JGR...10625301L} Leblanc, Y., Dulk, 
G.~A., Vourlidas, A., \& Bougeret, J.-L.\ 2001, \jgr, 106, 25301 


\bibitem[Lengyel-Frey et al.(1985)]{1985A&A...151..215L} Lengyel-Frey, D., 
Stone, R.~G., \& Bougeret, J.~L.\ 1985, \aap, 151, 215 

\bibitem[Lengyel-Frey \& Stone(1989)]{1989JGR....94..159L} Lengyel-Frey, 
D., \& Stone, R.~G.\ 1989, \jgr, 94, 159 


\bibitem[Malitson et al.(1973)]{1973ApJL...14..11} 
Malitson, H. H., Fainberg, J., \& Stone, R.G 1973, Astrophys. Lett., 14, 11 

\bibitem[Manchester et al. 2005]{Man05} Manchester, W. B., Gombosi, D., L., De Zeeuw, D. L., Sokolov, I. V., Roussev, I. I., Powell, K. G., Kota, J., Toth, G., \& Zurbuchen, T. H. 2005, \apj 622, 1225

\bibitem[Mann et al.(1999)]{1999Ap&SS.264..489M} Mann, G., Classen, H.-T., 
Motschmann, U., Kunow, H., \& Dr{\"o}ge, W.\ 1999, \apss, 264, 489 

\bibitem[Mann et al.(2001)]{2001JGR...10625323M} Mann, G., Classen, H.-T., 
\& Motschmann, U.\ 2001, \jgr, 106, 25323 


\bibitem[Mewaldt et al.(2005)]{2005JGRA..11009S18M} Mewaldt, R.~A., et al.\ 
2005, Journal of Geophysical Research (Space Physics), 110, 9 


\bibitem[Nelson \& Melrose(1985)]{1985srph.book..333N} Nelson, G.~J., \& 
Melrose, D.~B.\ 1985, Solar Radiophysics: Studies of Emission from the Sun 
at Metre Wavelengths, 333 

\bibitem[Newkirk, G. (1967)]{New67} Newkirk, G. A., Jr. 1967, ARAA 5, 213

\bibitem[Ramaty(1969)]{Ramaty_1969}  Ramaty, R. 1969, \apj, 158, 753

\bibitem[Ramaty \& Petrosian(1972)]{1972ApJ...178..241R} Ramaty, R., \& Petrosian, V.\ 1972, \apj, 178, 241

\bibitem[Reiner et al.(1997)]{1997cpsh.conf..183R} Reiner, M.~J., Kaiser, 
M.~L., Fainberg, J., Bougeret, J.-L., \& Stone, R.~G.\ 1997, ESA SP-415: 
Correlated Phenomena at the Sun, in the Heliosphere and in Geospace, 183 

\bibitem[Reiner et al.(1998)]{1998JGR...10329651R} Reiner, M.~J., Kaiser, 
M.~L., Fainberg, J., \& Stone, R.~G.\ 1998, \jgr, 103, 29651 

\bibitem[Reiner \& Kaiser(1999)]{1999JGR...10416979R} Reiner, M.~J., \& 
Kaiser, M.~L.\ 1999, \jgr, 104, 16979 

\bibitem[Reiner et al.(2000)]{2000ApJ...529L..53R} Reiner, M.~J., Kaiser, 
M.~L., Plunkett, S.~P., Prestage, N.~P., \& Manning, R.\ 2000, \apjl, 529, 
L53 

\bibitem[Reiner et al.(2001)]{2001JGR...10629989R} Reiner, M.~J., Kaiser, 
M.~L., \& Bougeret, J.-L.\ 2001, \jgr, 106, 29989 

\bibitem[Reiner et al.(2003)]{2003ApJ...590..533R} Reiner, M.~J., 
Vourlidas, A., Cyr, O.~C.~S., Burkepile, J.~T., Howard, R.~A., Kaiser, 
M.~L., Prestage, N.~P., \& Bougeret, J.-L.\ 2003, \apj, 590, 533 

\bibitem[Reiner et al.(2006)]{2006SoPh..234..301R} Reiner, M.~J., Kaiser, 
M.~L., Fainberg, J., \& Bougeret, J.-L.\ 2006, \solphys, 234, 301 

\bibitem[Reynolds, S. P., \& Chevalier, R. A. (1981)]{sc81} Reynolds, S. P., \& Chevalier, R. A. 1981, \apj 245, 912

\bibitem[Saito, K. (1970)]{Sai70} Saito, K. 1970, Ann. Tokyo Astron. Obs., Ser. 2, 12, 53

\bibitem[Saito, K., Poland, A. I., \& Munro, R. H. 1977]{Sai77} Saito, K., Poland, A. I., \& Munro, R. H. 1977, Sol. Phys. 55, 121


\bibitem[Simnett(2005)]{2005JGRA..11009S01S} Simnett, G.~M.\ 2005, Journal 
of Geophysical Research (Space Physics), 110, 9 

\bibitem[Simnett(2006)]{2006SoPh..237..383S} Simnett, G.~M.\ 2006, 
\solphys, 237, 383 


\bibitem[Smerd \& Dulk(1971)]{1971IAUS...43..616S} Smerd, S.~F., \& Dulk, 
G.~A.\ 1971, Solar Magnetic Fields, 43, 616 

\bibitem[Stewart(1985)]{1985srph.book..361S} Stewart, R.~T.\ 1985, Solar 
Radiophysics: Studies of Emission from the Sun at Metre Wavelengths, 361 

\bibitem[Uchida(1974)]{1974SoPh...39..431U} Uchida, Y.\ 1974, \solphys, 39, 
431 

\bibitem[Vr{\v s}nak et al.(2001)]{2001A&A...377..321V} Vr{\v s}nak, B., 
Aurass, H., Magdaleni{\'c}, J., \& Gopalswamy, N.\ 2001, \aap, 377, 321 

\bibitem[Vr{\v s}nak et al.(2006)]{2006A&A...448..739V} Vr{\v s}nak, B., 
Warmuth, A., Temmer, M., Veronig, A., Magdaleni{\'c}, J., Hillaris, A., \& 
Karlick{\'y}, M.\ 2006, \aap, 448, 739 

\bibitem[Wagner \& MacQueen(1983)]{1983A&A...120..136W} Wagner, W.~J., \& 
MacQueen, R.~M.\ 1983, \aap, 120, 136 

\bibitem[Wang et al. 2005]{Wang05} Wang, C., Du, D., \& Richardson, J. D. 2005, \jgr 110, A10107

\bibitem[Wild(1950)]{1950AuSRA...3..399W} Wild, J.~P.\ 1950, Australian 
Journal of Scientific Research A Physical Sciences, 3, 399 

\bibitem[Woo et al.(1995)]{1995GeoRL..22..329W} Woo, R., Armstrong, J.~W., 
Bird, M.~K., \& Patzold, M.\ 1995, \grl, 22, 329 



\end{thebibliography}
\end{document}